\documentclass{webofc}
\usepackage[varg]{txfonts}   
\usepackage{listings}
\usepackage{hyperref}
\usepackage{wrapfig}
\usepackage{gensymb}
\begin{document}
\title{Inference for Stellar Opacities from Seismic Studies of the Hybrid $\beta$ Cep/SPB pulsators}
%
%

\author{\firstname{Przemys{\l}aw} \lastname{Walczak}\inst{1}\fnsep\thanks{\href{mailto:walczak@astro.uni.wroc.pl}{\tt walczak@astro.uni.wroc.pl}} \and
        \firstname{Jadwiga} \lastname{Daszy{\'n}ska-Daszkiewicz}\inst{1}
        \and
        \firstname{Alexey} \lastname{Pamyatnykh}\inst{1}
}

\institute{Astronomical Institute Wroc{\l}aw University, ul. Kopernika 11, 51-622 Wrocław, Poland
\and
           Nicolaus Copernicus Astronomical Center Polish Academy of Sciences, ul. Bartycka 18, 00-716 Warsaw, Poland
          }

\abstract{%
We present a comprehensive seismic study of the three pulsating stars of $\beta$ Cep/SPB type: $\nu$ Eridani, 12 Lacertae and $\gamma$ Pegasi. Models with the modified mean opacity profile are constructed in order to account for both the observed frequency range and the values of some individual frequencies. To decrease the number of possible solutions, we make use of the non-adiabatic parameter $f$, whose value is very sensitive to subphotospheric layers where pulsations are driven. This complex seismic modelling show the need for a significant modification of the opacity profile.
}
\maketitle

\section{Introduction}\label{sec:intro}

The hybrid pulsators, exhibiting oscillations both in low order p/g modes and high
order g modes, pose a challenge for pulsation theory because the range of their observed frequencies cannot be explained by standard models. To overcome this problem an increase of the opacity at the certain depths was postulated by many authors (i.e. \cite{2004MNRAS.350.1022P}, \cite{2012MNRAS.422.3460S}, \cite{2016MNRAS.455L..67M}). This approach is justified by recent experiments \cite{Bailey_e2015}, where higher than predicted opacities of iron and nickel were measured.

Here we analyse the three well known hybrid $\beta$ Cep/SPB pulsators: $\nu$ Eri (HD 29248), $\gamma$ Peg (HD 886) and 12 Lac (HD 214993). The stars pulsate in many modes of different types, i.e. in low frequency high order g modes (SPB-type) and high frequency low order p/g-modes ($\beta$ Cep-type).

In Sect.\,\ref{sec:sec-1}, we summarize the basic information about the stars. In Sect.\,\ref{sec:sec-2}, we search the pulsational models with the standard and modified opacity profile.
Conclusions end the paper.

\section{The three analysed pulsators}\label{sec:sec-1}


All the three pulsators are massive stars of the B2 spectral type. The effective temperature of $\nu$ Eri amounts to $\log{T_{\rm{eff}}}=4.345\pm0.014$, as determined by \cite{JDD2005}. $\gamma$ Peg is the coolest star in our sample with $\log{T_{\rm{eff}}}=4.316\pm0.017$ \citep{2012A&A...538A.143K}. For 12 Lac we used $\log{T_{\rm{eff}}}=4.375\pm0.018$ \cite{2006MNRAS.365..327H}. With the parallaxes by \cite{2007A&A...474..653V} we derived luminosities $\log{L/L_{\odot}}=3.807\pm0.045$, $\log{L/L_{\odot}}=3.710\pm0.065$ and $\log{L/L_{\odot}}=4.077\pm0.11$ for $\nu$ Eri, $\gamma$ Peg and 12 Lac, respectively.

The position of the three target stars on the Hertzsprung-Russell diagram is shown in Fig.\,\ref{fig:puls-models}. The evolutionary tracks were calculated with the metallicity parameter $Z=0.015$, initial hydrogen abundance $X=0.7$ and with no overshooting from the convective core. We assumed the solar chemical
\begin{wrapfigure}{r}{0.49\textwidth}
  \centering
  \includegraphics[width=0.48\textwidth, clip]{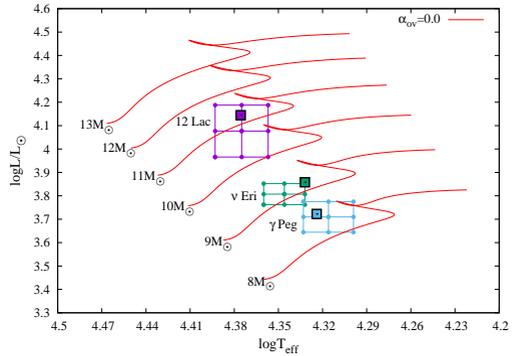}
  \caption{The observational error boxes of $\nu$ Eri, $\gamma$ Peg and 12 Lac in the  Hertzsprung-Russell diagram. The evolutionary tracks were computed for metallicity $Z = 0.015$ and the overshooting parameter $\alpha_{\rm{ov}}=0$. Models marked as squares are described in the text.}
  \label{fig:HR}
\end{wrapfigure}
composition by \cite{AGSS09} and the new opacities from Los Alamos National Laboratory, OPLIB \cite{OPLIB2,OPLIB1}. Evolutionary calculations were performed using the Warsaw-New Jersey code \cite{1998A&A...333..141P}. All stars seems to be well on the main--sequence evolution phase. The models marked with squares are explained in Sect.\,\ref{sec:sec-2}.

$\nu$ Eri was identified as the first early B-type hybrid pulsator. The analysis of the data gathered during extensive 2003-2005 world-wide observational campaigns \cite{Handler2004,J05,Aerts2004,deRidder2004} revealed 14 pulsational frequencies. Two of them were low frequency modes and 12 - high frequency modes.
Recently, the star has been observed by the BRITE-Constellation \cite{BRITE1}. \cite{Handler2017} derived 17 pulsational frequencies including 7 low frequency modes and 10 high frequency modes. $\gamma$ Peg has been observed by the MOST satellite \citep{2003PASP..115.1023W}. The analysis of the data revealed 14 pulsational frequencies: 8 in the high frequency range and 6 in the low frequency range \citep{2009ApJ...698L..56H}. 12 Lac was the aim of the multisite observational campaign carried out in the years 2003-2004. \cite{2006MNRAS.365..327H} reported 11 pulsational frequencies derived form the light curves (one of them was a low frequency mode and 10 - high frequency modes). The oscillation spectra of the three puslators are shown in Fig.\,\ref{fig:puls-models}.

\section{Pulsational Models}\label{sec:sec-2}

In order to perform instability analysis, we calculated non-adiabatic pulsational models using the code written by Dziembowski \cite{1977AcA....27...95D}.
The computations were based on the three opacity sources: OPAL \cite{OPAL}, OP \cite{OP} and OPLIB.

\begin{figure}[h]
\centering
\includegraphics[width=0.49\textwidth]{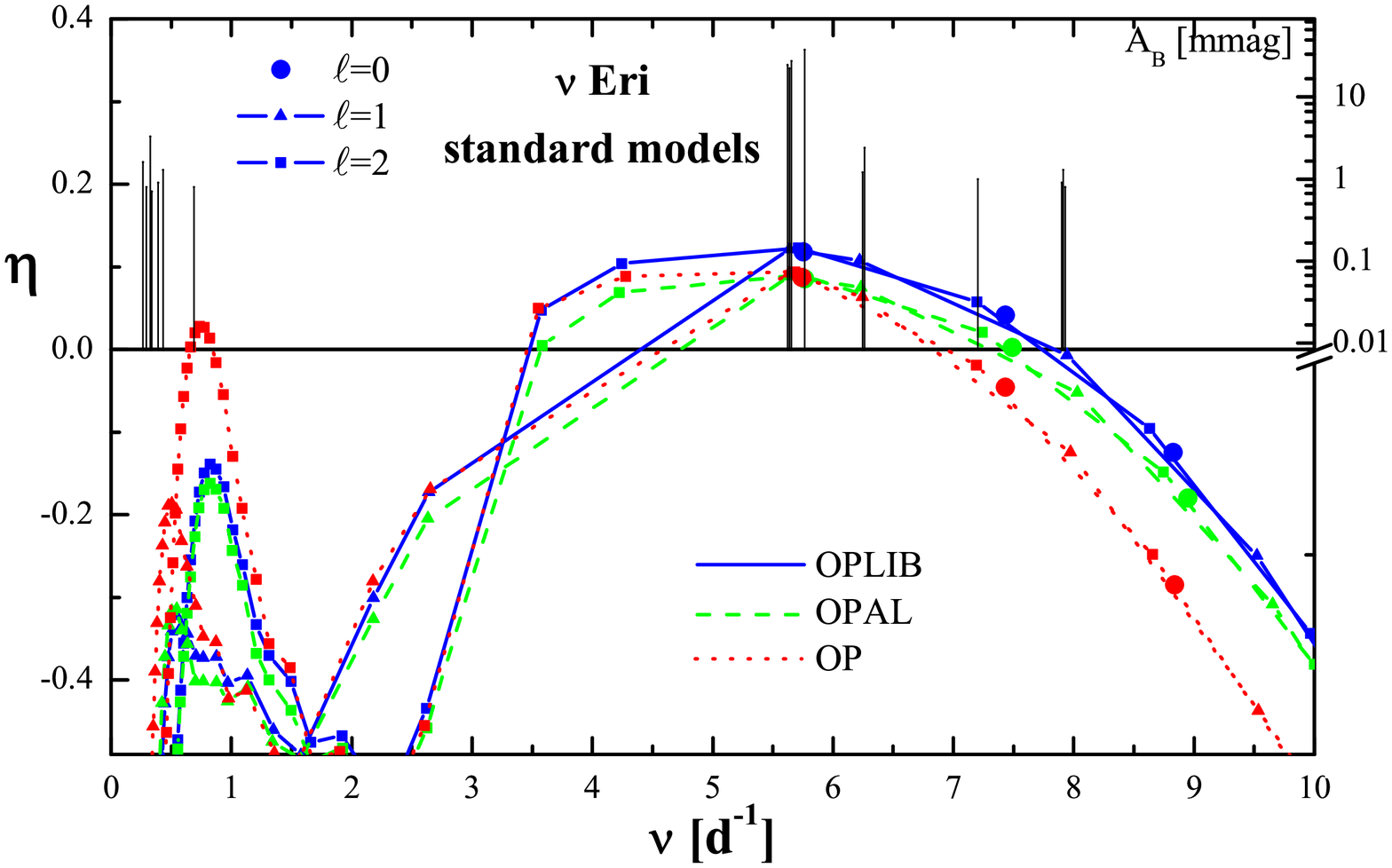}
\includegraphics[width=0.49\textwidth]{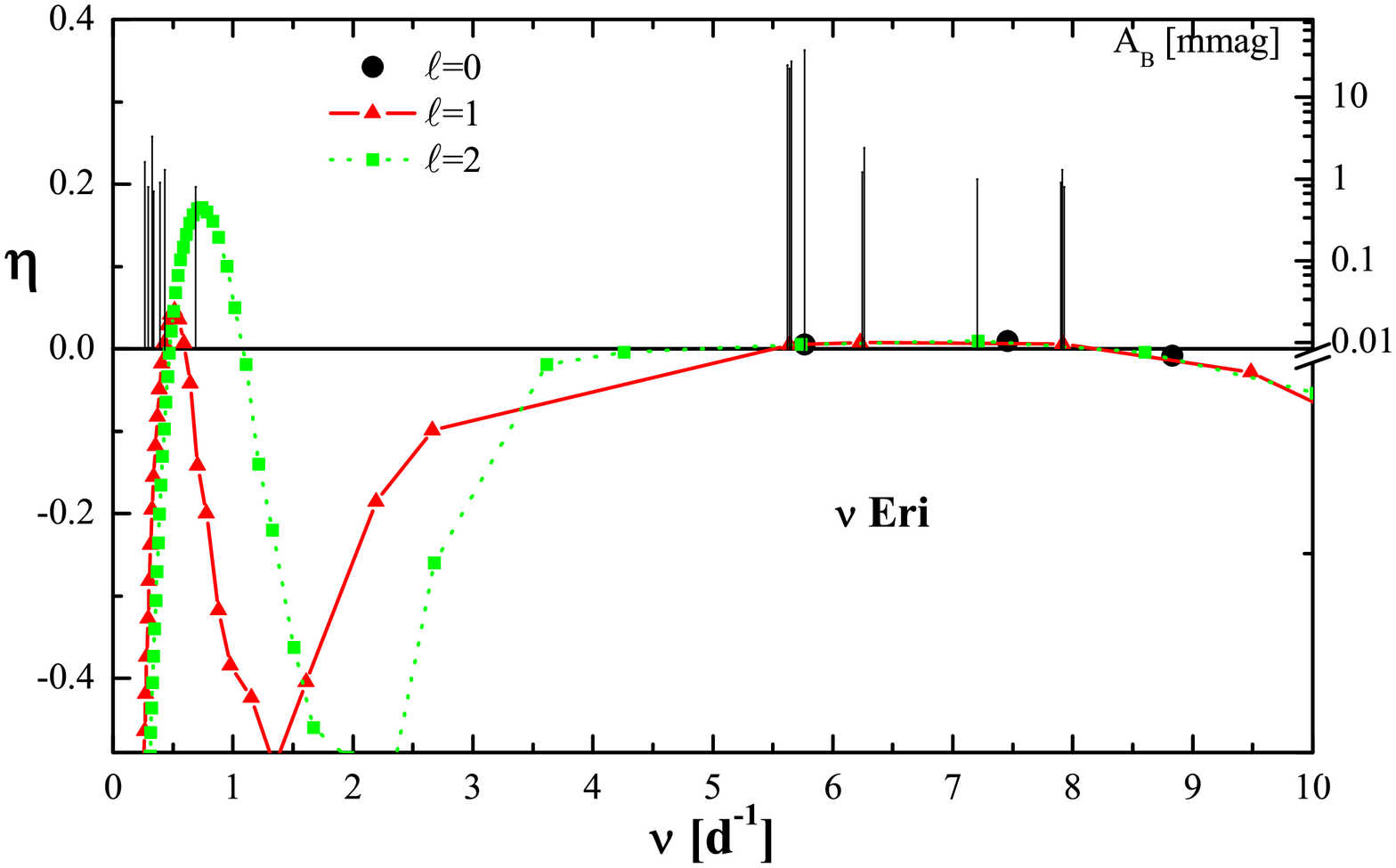}
\includegraphics[width=0.49\textwidth]{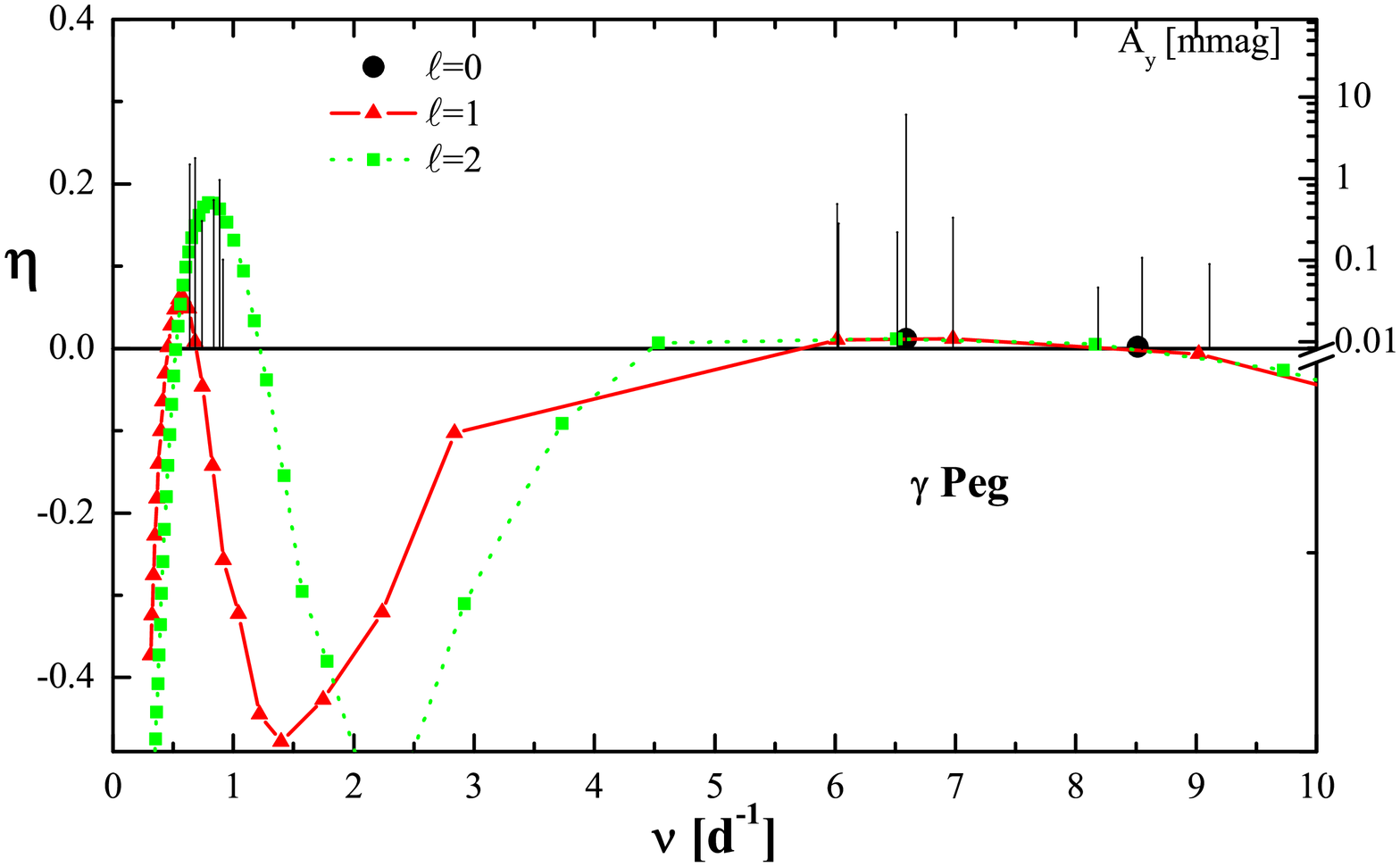}
\includegraphics[width=0.49\textwidth]{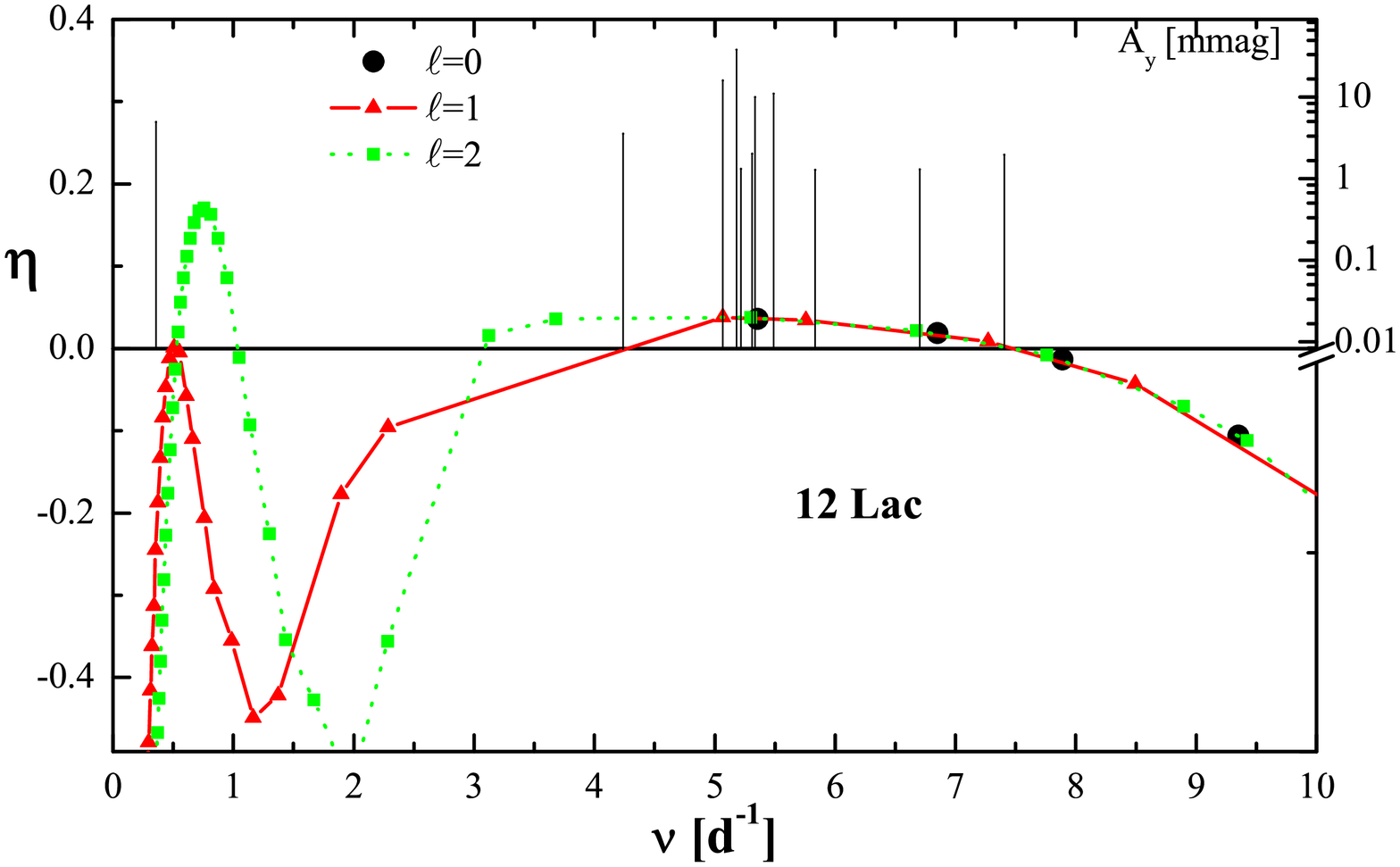}
\caption{The instability parameter, $\eta$, as a function of frequency for modes with the degree $\ell=0-2$. In the upper-left panel, we showed the standard models of $\nu$ Eri calculated with three opacity data. In the upper-right panel, the best model calculated with the modified OPLIB opacities is shown. In the bottom panels, we plot the modified OPLIB models of $\gamma$ Peg and 12 Lac, respectively.  The opacity modification parameters are given in Table 1. The vertical lines indicate the observed frequencies.}
\label{fig:puls-models}
\end{figure}

In the case of $\nu$ Eri, we selected models, that fit the three frequencies, the radial mode (5.76 d$^{-1}$) and two centroids of the dipole modes (5.63 d$^{-1}$ and 6.24 d$^{-1}$). An example of such models calculated with the three opacity data are plotted in the upper left panel of Fig.\,\ref{fig:puls-models}, where we showed the instability parameter, $\eta$, as a function of the frequency, $\nu$. A mode is excited if $\eta>0$. We marked the observed frequencies of $\nu$ Eri with vertical lines. Their height indicates the amplitude in the BRITE Blue filter. We assumed metallicity $Z=0.015$, mass $M=9.5M_{\odot}$ and the overshooting parameter, $\alpha_{\text{ov}}$, from 0.07 up to 0.09. The value of $\alpha_{\rm{ov}}$ was changed to fit the frequencies of the dipole mode centroids.

We see, that neither OPAL nor OP nor OPLIB models can explain the whole observed frequency range. The OPLIB model has unstable  high frequency modes, but low frequency modes are not excited. The OPAL model is very similar to the OPLIB model with slightly smaller instabilities, especially for high frequencies. The OP model has the highest instability in the low frequency region but the instability of p modes at the highest frequencies ($\nu\gtrsim6.5$ d$^{-1}$) is much reduced. Changing the model parameters in a reasonable range do not solve the problem.
The results for $\nu$ Eri have been already published in \citep{2017MNRAS.466.2284D} and here we give only a summary and brief outline of the approach.

In order to increase the excitation of modes, we changed artificially the standard mean opacity profile according to the formula:
\begin{equation}
  \kappa (T)=\kappa_0(T) \left[1+\sum_{i=1}^N b_i \cdot\exp\left( -\frac{(\log T-\log T_{0,i})^2}{a_i^2}\right) \right],
  \label{eq:eq-1}
\end{equation}
where $\kappa_0(T)$ is the original opacity profile and $(a,~b,~T_0)$ are parameters of the Gaussian describing the width, height and a position of the maximum, respectively.

By changing the value of the parameters $a$, $b$ and $T_0$, we can find a lot of seismic models that are unstable in the regions of interest. To constrain the number of combinations of $a$, $b$ and $T_0$, we included the nonadiabatic parameter $f$ in our fitting. The parameter $f$ is defined as the ratio of the bolometric flux perturbation to the radial displacement at the level of the photosphere \cite{JDD2003,JDD2005} and its value is very sensitive to the opacity of the outer layers. The empirical values of $f$ can be derived from multicolor photometry and radial velocity measurements \citep{JDD2003,JDD2005}. They depend also on atmosphere models, therefore we used both LTE and non-LTE atmosphere models by \cite{Kurucz2004} and \cite{2007ApJS..169...83L}, respectively. Since the non-adiabatic effects of pulsation cause the phase shift of the flux changes, we need to consider both, the amplitude, $|f|$, and the phase lag, $\Psi=arg(f)-180^{\circ}$. The value of $\Psi$ describes the phase shift between the maximum of the flux and the minimum of the radius.

Finding the models that fit simultaneously the instability range, the values of the frequencies and the value of $f$ for the radial mode was a challenging task. We searched an extensive grid of models changing parameters of eq.\,\ref{eq:eq-1} with steps: $\Delta a=0.001$, $\Delta b=0.05$ and $\Delta T_0=0.005$. The $\log{T_0}$ parameter was chosen from the range 5.0-5.5. It turned out, that only models with quite complicated opacity modification meet all requirements.

Our best model calculated with the OPLIB opacity modification is shown in the upper right panel of Fig.\,\ref{fig:puls-models}, where we plotted the instability parameter as a function of the mode frequency. The position of the model on the HR diagram is marked as a big square nearly inside the $\nu$ Eri error box in Fig.\,\ref{fig:HR}. The model parameters are as follows: $M=9.0M_{\odot}$, $\log{T_{\text{eff}}}=4.332$, $\log{L/L_{\odot}}=3.858$, $Z=0.015$ and $\alpha_{\text{ov}}=0.163$.
The parameters of the opacity modifications are given in the first line of Table~\ref{tab:tab-1}.
The mean opacity has been changed at the three depths, $\log{T_{0,1}}=5.06$ (corresponding to the opacity bump identified by \cite{Cugier2012,Cugier2014}), $\log{T_{0,2}}=5.22$ (corresponding to the maximum contribution of chromium and manganese, \cite{2012MNRAS.422.3460S}) and $\log{T_{0,3}}=5.46$ (corresponding to the maximum contribution of nickel, \cite{2012MNRAS.422.3460S}).
This $\kappa-$modified model can account for all observed high frequency modes. The lowest observed frequency modes are not excited, but the instability parameter in this frequency region is significantly higher than in standard models. Moreover, most of the peaks can be explained by the rotational splitting.

We performed the same analysis for $\gamma$ Peg and 12 Lac and we found that in these cases a very similar opacity modification works as well. In case of $\gamma$ Peg we calculated models that fit two frequencies: the radial mode, $\nu=6.59$ d$^{-1}$, and the dipole mode, $\nu=6.02$ d$^{-1}$. Also the modes fitting the two frequencies were constructed for 12 Lac: the radial mode, $\nu=5.33$ d$^{-1}$, and the dipole mode, $\nu=5.07$ d$^{-1}$. The instability parameters of the best models of $\gamma$ Peg and 12 Lac are shown in the bottom left and right panels of Fig.\,\ref{fig:puls-models}, respectively. The vertical lines correspond to the observed frequencies and their height indicates the amplitude in the Str\"omgrem $y$ filter. For $\gamma$ Peg, the model parameters are: $M=8.06M_{\odot}$, $\log{T_{\text{eff}}}=4.324$, $\log{L/L_{\odot}}=3.723$, $Z=0.012$ and $\alpha_{\text{ov}}=0.262$. The model of 12 Lac has the parameters: $M=11.152M_{\odot}$, $\log{T_{\text{eff}}}=4.376$, $\log{L/L_{\odot}}=4.145$, $Z=0.015$ and $\alpha_{\text{ov}}=0.1$. The parameters of the opacity modifications are given in second and third lines of Table~\ref{tab:tab-1}.

The model of $\gamma$ Peg has a high instability in the low frequency range and all observed SPB-type modes are explained by this model. In the case of high frequencies only one mode is slightly stable, ie. that with the highest frequency. For 12 Lac, the modified model has unstable high frequency modes, but the only low frequency mode is stable. Nevertheless, the instability parameter is quite high in the SPB-type mode region for all three stars.

The empirical values of $f$ for $\nu$ Eri and 12 Lac are inside of the empirical uncertainties. In case of $\gamma$ Peg, the amplitude $|f|$ is reproduced within $1\sigma$ errors, while the phase lag, $\Psi$, within $2\sigma$ errors.

\begin{table}[h]
\centering
\caption{Parameters of the opacity modification for models described in the text.}
\begin{tabular}{cccccccccc}
\hline
star         & $\log{T_{0,1}}$ & $a_1$ &$b_1$&$\log{T_{0,2}}$&$a_2$&$b_2$ &$\log{T_{0,3}}$ &$a_3$& $b_3$ \\ \hline
$\nu$ Eri    & 5.06 & 0.500 & -0.54 &5.22 &0.088 &+0.34 &5.46 & 0.061 & +1.80 \\
$\gamma$ Peg & 5.06 & 0.447 & -0.60 &5.22 &0.067 &+0.50 &5.46 & 0.063 & +2.10\\
12 Lac       & 5.06 & 0.447 & -0.25 &5.22 &0.082 &+0.50 &5.46 & 0.077 & +2.00\\\hline
\end{tabular}
\label{tab:tab-1}       
\end{table}

\section{Conclusions}\label{sec:con}

The standard asteroseismic models of the hybrid $\beta$ Cep/SPB pulsators cannot explained the observed properties.
The biggest problem is the theoretical instability range, which is usually too small in comparison with the observations.
As a possible solution, we proposed a modification of stellar opacities at certain depths. It goes in line with
the recent experimental results which give the higher opacities for iron and nickel.
The aim was to reproduce the observed frequency range, the values of some individual frequencies and the parameter $f$ which is very sensitive to the opacity changes
in the subphotospheric layers. Such complex seismic modelling reduces greatly the number of possible opacity modification and make our modelling more justifiable.

The most interesting and striking result is  than the modification of the opacity profile of our best model is very similar for all the three analysed stars. In each case, the opacity modifications are necessary at the three depths: an increase by about 200\% at $\log{T_{0,3}}=5.470$, an increase by about 30-50\% at $\log{T_{0,2}}=5.220$ and a decrease by about 50\% in the uppermost layers at $\log{T_{0,1}}=5.065$.

The necessity of decreasing opacity near $\log{T}=5.065$, where a new maximum was identified by \citep{Cugier2012,Cugier2014}, is quite bothersome and we do not have an explanation for that. Also a huge increase of $\kappa$ at $\log{T}=5.470$ seems to be very surprising. This may results from both the inhomogeneous chemical composition and inaccuracies in the opacity calculations. Intensive further studies are needed to resolve this issue.

\begin{acknowledgement}
\noindent\vskip 0.2cm
\noindent {\em Acknowledgments}: This work was financially supported by the Polish NCN grant 2015/17/B/ST9/02082. Calculations have been partly carried out using resources provided by Wroc{\l}aw Centre for Networking and Supercomputing (http://www.wcss.pl), grant No. 265. 
\end{acknowledgement}

%

\bibliography{PWBiblio}

\begin{thebibliography}{29}

\bibitem{2004MNRAS.350.1022P}
A.A. {Pamyatnykh}, G.~{Handler}, W.A. {Dziembowski}, MNRAS \textbf{350}, 1022
  (2004)

\bibitem{2012MNRAS.422.3460S}
S.~{Salmon}, J.~{Montalb{\'a}n}, T.~{Morel}, {et al.}, MNRAS \textbf{422}, 3460
  (2012)

\bibitem{2016MNRAS.455L..67M}
E.~{Moravveji}, MNRAS \textbf{455}, L67 (2016)

\bibitem{Bailey_e2015}
J.E. {Bailey}, T.~{Nagayama}, G.P. {Loisel}, {et al.}, Nature \textbf{517}, 56
  (2015)

\bibitem{JDD2005}
J.~{Daszy{\'n}ska-Daszkiewicz}, W.A. {Dziembowski}, A.A. {Pamyatnykh}, A\&A
  \textbf{441}, 641 (2005)

\bibitem{2012A&A...538A.143K}
M.~{Koleva}, A.~{Vazdekis}, A\&A \textbf{538}, A143 (2012)

\bibitem{2006MNRAS.365..327H}
G.~{Handler}, M.~{Jerzykiewicz}, E.~{Rodr{\'{\i}}guez}, {et al.}, MNRAS
  \textbf{365}, 327 (2006)

\bibitem{2007A&A...474..653V}
F.~{van Leeuwen}, A\&A \textbf{474}, 653 (2007)

\bibitem{AGSS09}
M.~{Asplund}, N.~{Grevesse}, A.J. {Sauval}, P.~{Scott}, ARAA \textbf{47}, 481
  (2009)

\bibitem{OPLIB2}
J.~{Colgan}, D.P. {Kilcrease}, N.H. {Magee}, {et al.}, High Energy Density
  Physics \textbf{14}, 33 (2015)

\bibitem{OPLIB1}
J.~{Colgan}, D.P. {Kilcrease}, N.H. {Magee}, {et al.}, ApJ \textbf{817}, 116
  (2016)

\bibitem{1998A&A...333..141P}
A.A. {Pamyatnykh}, W.A. {Dziembowski}, G.~{Handler}, H.~{Pikall}, A\&A
  \textbf{333}, 141 (1998)

\bibitem{Handler2004}
G.~{Handler}, R.R. {Shobbrook}, M.~{Jerzykiewicz}, {et al.}, MNRAS
  \textbf{347}, 454 (2004)

\bibitem{J05}
M.~{Jerzykiewicz}, G.~{Handler}, R.R. {Shobbrook}, {et al.}, MNRAS
  \textbf{360}, 619 (2005)

\bibitem{Aerts2004}
C.~{Aerts}, P.~{De Cat}, G.~{Handler}, {et al.}, MNRAS \textbf{347}, 463 (2004)

\bibitem{deRidder2004}
J.~{De Ridder}, J.H. {Telting}, L.A. {Balona}, {et al.}, MNRAS \textbf{351},
  324 (2004)

\bibitem{BRITE1}
W.W. {Weiss}, S.M. {Ruci\'nski}, A.F.J. {Moffat}, {et al.}, PASP \textbf{126},
  573 (2014)

\bibitem{Handler2017}
G.~{Handler}, M.~{Rybicka}, A.~{Popowicz}, {et al.}, MNRAS \textbf{464}, 2249
  (2017)

\bibitem{2003PASP..115.1023W}
G.~{Walker}, J.~{Matthews}, R.~{Kuschnig}, {et al.}, PASP \textbf{115}, 1023
  (2003)

\bibitem{2009ApJ...698L..56H}
G.~{Handler}, J.M. {Matthews}, J.A. {Eaton}, {et al.}, ApJ \textbf{698}, L56
  (2009)

\bibitem{1977AcA....27...95D}
W.~{Dziembowski}, AcA \textbf{27}, 95 (1977)

\bibitem{OPAL}
C.A. {Iglesias}, F.J. {Rogers}, ApJ \textbf{464}, 943 (1996)

\bibitem{OP}
M.J. {Seaton}, MNRAS \textbf{362}, L1 (2005)

\bibitem{2017MNRAS.466.2284D}
J.~{Daszy{\'n}ska-Daszkiewicz}, A.A. {Pamyatnykh}, P.~{Walczak}, {et al.},
  MNRAS \textbf{466}, 2284 (2017)

\bibitem{JDD2003}
J.~{Daszy{\'n}ska-Daszkiewicz}, W.A. {Dziembowski}, A.A. {Pamyatnykh}, A\&A
  \textbf{407}, 999 (2003)

\bibitem{Kurucz2004}
R.~{Kurucz}, http:// kurucz.harvard.edu  (2004)

\bibitem{2007ApJS..169...83L}
T.~{Lanz}, I.~{Hubeny}, ApJ \textbf{169}, 83 (2007)

\bibitem{Cugier2012}
H.~{Cugier}, A\&A \textbf{547}, 42 (2012)

\bibitem{Cugier2014}
H.~{Cugier}, A\&A \textbf{565}, 76 (2014)

\end{thebibliography}

\end{document}